\documentclass[a4paper]{article}
\usepackage{amsmath}	
\usepackage{graphicx}

\makeatletter

\@addtoreset{equation}{section}
\makeatother

\newcommand{\Tr}{\mathop{\rm Tr}\nolimits}
\newcommand{\psibar}{\overline{\psi}}

\begin{document}

\begin{flushright}
DFTT 06/2010\\
June, 2010\\
\end{flushright}
\bigskip\bigskip

\begin{center}
\renewcommand{\thefootnote}{\fnsymbol{footnote}}
\newcommand{\inst}[1]{\mbox{$^{\text{\textnormal{#1}}}$}}
{\LARGE
 A Method for Measuring the Witten Index\\
 Using Lattice Simulation
}\\[8ex]
%
{\large
Issaku Kanamori\footnote{\texttt{kanamori@to.infn.it}}
}\\
%
{\large\itshape
INFN sezione di Torino, and\\
Dipartimento di Fisica Teorica,
Universita di Torino\\
I-10125 Torino, Italy\\[3ex]
}
\end{center}
\bigskip\bigskip
\setcounter{footnote}{0}

\begin{abstract}
 We propose a method to measure the Witten index using lattice
 simulation.  A requirement for the lattice model is that it has at least
 one exact supersymmetry at finite lattice spacing.
 We prove the validity of the method in case of the supersymmetric
 quantum mechanics, where the index is well known.
\end{abstract}

\section{Introduction}

Recent developments on lattice formulations of supersymmetric
theories\footnote{For a recent review, see Ref.~\cite{Catterall:2009it}.
}
make it possible to perform numerical simulations to study non-perturbative aspects 
of the theories.
The most important feature in
these developments is that the action has (at least) one exact
fermionic symmetry at finite lattice spacing.
This was pioneered 
in~\cite{Kaplan:2002wv} for gauge theories and
in~\cite{Catterall:2001fr} for non-gauge theories.
It turned out later that this exact symmetry is 
the scalar part of supersymmetry in terms of the topological twist.  
The symmetry is strong enough to guarantee an automatic 
restoration of the remaining part of the supersymmetry without
any fine tuning for the 1- or 2-dimensional case.
This restoration is manifestly confirmed numerically in 2-dimensional
$\mathcal{N}=(2,2)$ super Yang-Mills \cite{Kanamori:2008bk} using a
model by F.~Sugino~\cite{Sugino:2004qd}, which also has one
 exact supersymmetry at finite lattice spacing.
A more ambitious approach which tries to keep all of the supersymmetry was
also proposed by D'Adda et.al. 
in \cite{sapporo-torino:link}
and its Hopf algebraic
structure is studied in~\cite{sapporo-torino:hopf}.
A new proposal along this line is found in~\cite{DFKKS2010}.
However, without introducing non-locality and/or extra structure
such as noncommutativity,
only a part of supersymmetry out of full supersymmetry
can be realized on the lattice;
Especially, the equivalence of~\cite{sapporo-torino:link} to other
approaches is well studied in~\cite{Damgaard:2007eh}.
Not only simple spacetime lattice
regularization but also regularization 
in the momentum space \cite{Hanada:2007ti}
sheds some light on how to regularize supersymmetric
theories (see also \cite{Kadoh:2009sp, Bergner:2009vg}).  
A different approach based on the large-$N$ limit is
proposed in~\cite{PWMM}.
Further developments in the formulation on the lattice are found in~\cite{
Kadoh:2009yf,Kadoh:2009rw,Hanada:2010kt,Kawai:2010yj}.

One of the most interesting non-perturbative aspects of supersymmetry
is the spontaneous breaking of the symmetry.  
Supersymmetry is broken in our current
universe and the breaking should be the result of a non-perturbative effect,
unless it is broken from the beginning at the tree level.
To study the breaking of supersymmetry using lattice simulation, 
a Hamiltonian approach was used in \cite{Beccaria:2004pa}.
The present author with H.~Suzuki and F.~Sugino proposed a method for detecting
the spontaneous supersymmetry breaking using lattice simulation
in~\cite{KSS07}, where
they measured the vacuum energy. They also pointed out a possible relation 
between spontaneous supersymmetry breaking and the existence of a sign
problem in the simulation, which comes from the phase of 
the Pfaffian of the fermion bilinear operator.

Another line of application with large-$N$ limit 
is found in the gauge/gravity duality.
In particular, extensive studies of 1-dimensional model for this purpose 
are found in Refs.~\cite{AHNTetal, Catterall-Wiseman}.
The large-$N$ property of 2-dimensional $\mathcal{N}=(2,2)$ system has
been studied with the same motivation \cite{Hanada:2009hq}, although
this system has smaller number of supersymmetries.

In this paper, we propose another method of detecting the spontaneous
breaking of supersymmetry.  We measure the Witten index~\cite{Witten:1982df}.
If the index is not vanishing, there exists at least one supersymmetric vacuum
and supersymmetry is not broken.  If the index is vanishing, 
supersymmetry may or may not be broken.
Since the Witten index is a partition function 
with periodic boundary conditions in the time
direction~\cite{Cecotti:1981fu,Fujikawa:1982nt}, 
it is of crucial importance to
determine the normalization factor in simulation.
Usually what we obtain in the simulation are expectation values
normalized by the partition function, but what we need here is the
normalization itself.\footnote{
An interesting simulation method which allows to determine 
the normalization and thus the index is proposed \cite{Kawai:2010yj} 
on the basis of the Nicolai map.
In a $0$-dimensional system, Monte Carlo integration was used 
to calculate the partition function in~\cite{Krauth:1998xh}.
}
Our idea to determine the factor consists of two ingredients.
First, we regard the derivation of the path integral formalism from the
operator formalism as a lattice regularization of the time coordinate.
This gives a correct normalization of the path integral measure.
Second, we measure a suitably chosen quantity, which cancels a
distribution functional $e^{-S}$ in the path integral.
Because of this cancellation, we can separate the contribution from
the normalization factor.
We use supersymmetric quantum mechanics (of $\mathcal{N}=2$ Wess-Zumino type)
\cite{Witten:1981nf} as an arena for checking our method.
This system can be formulated on the lattice in various ways\footnote{
Recent proposals with lattice or non-lattice are found in \cite{Hanada:2007ti,
Arianos:2008ai,
Kadoh:2009sp,Bergner:2009vg, DFKKS2010}.
See also \cite{Bergner:2007pu}.}.
There is a formulation which keeps a nilpotent supercharge $Q$ 
at finite lattice spacing, with which the action is written in 
a $Q$-exact form~\cite{Catterall:2000rv,Catterall:2003wd,Giedt:2004vb}. (See
also \cite{Beccaria:1998vi}.)

In supersymmetric Yang-Mills quantum mechanics with maximal
supersymmetry, 
which would be an interesting
application of our method, the index is of crucial importance.  
This system is a candidate for M-theory \cite{Banks:1996vh}, 
and to guarantee a suitable limit
which gives supergravity, the index should be one.
Related calculations are found in \cite{%
Yi:1997eg,Sethi:1997pa,Moore:1998et}.  
For this system, a numerical treatment in the Fock space is proposed \cite{
Wosiek:2002nm,Campostrini:2002mr,Campostrini:2004bs,Korcyl:2009iz}
which is also useful
to obtain the Witten index as well.
To the author's best knowledge, however,
only a little is known about the index from direct calculations of
the quantum mechanics so far.

In the next section, we describe our idea in detail.
Then we check the validity of the measure by analytically calculating
the index for supersymmetric quantum mechanics in the free case in
section~\ref{sec:analytical}.
In section~\ref{sec:numerical} , we numerically calculate the index of the
supersymmetric quantum mechanics 
and demonstrate that our method in fact reproduces the known index.

\section{Basic Idea}
\label{sec:idea}

The idea for measuring the Witten index is made of two ingredients.
One is for the regularization of the path integral measure
and the other is for the regularization of the integral.

Let us start with the quantum mechanics.
The standard way  to obtain the path integral formulation from the operator
formulation is discretizing  the temporal direction and then inserting a
complete set at each of the time slice:
\begin{equation}
 \langle q_{\rm fin} |e^{-i\hat{H}T} | q_{\rm ini} \rangle
  =  \left( \prod_{k=1}^{N-1}\int_{-\infty}^{\infty} dq_k \right)
      \langle q_{\rm fin} | e^{-ia\hat{H}} |q_{N-1} \rangle
      \langle q_{N-1} | e^{-ia\hat{H}} |q_{N-2} \rangle
      \cdots
      \langle q_{1} | e^{-ia\hat{H}} |q_{\rm ini} \rangle,
\end{equation}
where $|q_i\rangle$ is a \emph{normalized} state, $\hat{H}$ is the
Hamiltonian\footnote{
We explicitly denote the operators with hat ($\hat{\ }$) in this section.
}
of the system, and the time difference of 
the initial state $|q_{\rm ini}\rangle$
and the final state $|q_{\rm fin}\rangle$ is $T=aN$.
We regard this $a$ as the lattice spacing.  
Then the standard derivation\footnote{
For the sake of completeness, we give a brief review of the derivation
of path integral formulation in Appendix~\ref{app:pathintegral}.}
 gives the path integral measure
for a bosonic dimensionless lattice field $\phi^{\rm lat}$ as
follow:
\begin{equation}
  \int \mathcal{D}\phi^{\rm lat}
 =
 \prod_{k=0}^{N-1} \left[
 \left(\frac{1}{2\pi}\right)^{\frac{1}{2}}
 \int_{-\infty}^\infty d\phi_k^{\rm lat} \right].
 \label{eq:measure-boson}
\end{equation}
For dimensionless fermionic variables, we obtain
\begin{equation}
  \int \mathcal{D}\psi^*\mathcal{D}\psi 
  = \prod_{k=0}^{N-1} \int d\psi_k^*\,d\psi_k.
 \label{eq:measure-fermion}
\end{equation}
In this way, we can determine the natural measure for the path
integral.

In the lattice simulation, what we calculate is not the integration
discussed so far.  Instead, we calculate ensemble averages which give
the ratios of the integrations.  
We have to establish a relation between these averages and
the path integral regularized on the lattice.

In terms of the path integral, the expectation value is
\begin{equation}
 \langle A \rangle
  =\frac{\int \mathcal{D}\phi\, A\, e^{-S} }
   {\int \mathcal{D}\phi\,e^{-S}},
\end{equation}
where $S$ is the action of the system
and we omit fermions for a while. 
The normalization, that is, the
denominator is exactly the partition function which we are interested in.
Since it is made of a ratio and does not depend on the normalization of 
the path integral, it seems impossible to measure the partition function
with the correct normalization.
However, let us consider the following quantity:
\begin{equation}
 \langle e^{+S} e^{-\frac{1}{2}\sum_i\mu^2 (\phi_i^{\rm lat})^2} \rangle
 =\frac{\int \mathcal{D}\phi\, 
        \exp \left[-\frac{1}{2}\sum_i \mu^2 (\phi_i^{\rm lat})^2\right]}
   {\int \mathcal{D}\phi\, e^{-S} }
 \equiv
  \frac{C}{\int \mathcal{D}\phi\, e^{-S}},
 \label{eq:observable}
\end{equation}
where $\mu$ is an arbitrary dimensionless parameter.
With the measure defined in eq.~(\ref{eq:measure-boson}),
we obtain
\begin{equation}
 C= \int \mathcal{D}\phi\, e^{-\frac{1}{2}\sum_i\,  \mu^2 \phi_i^2}
  = \prod_{i=0}^{N-1} \left[  \left(\frac{1}{2\pi}\right)^\frac{1}{2}
      \int_{-\infty}^{\infty}d\phi_i\,
      e^{-\frac{1}{2} \sum_i\mu^2 \phi_i^2}
 \right]
  = \mu^{-N}.
 \label{eq:C}
\end{equation}
Here, we have renamed the dimensionless 
lattice field $\phi_i^{\rm lat} \to \phi_i$.
For any real value of $\mu$, we can determine $C$ analytically
which gives the overall normalization of the path integral.
Note that $C$ is given just as an integration defined in (\ref{eq:C}),
which is independent from the distribution of $\{\phi_i\}$.
Especially, we never use Gaussian distributed quantities.
The left hand side of eq.~(\ref{eq:observable}) 
is an observable in the simulation.
Therefore, we can obtain the value of the partition function through the
following expression:
\begin{equation}
 Z=\int \mathcal{D}\phi\, e^{-S}
  =\frac{C}{\langle e^{+S} e^{-\frac{1}{2}\sum_i  \mu^2 \phi_i^2} \rangle}.
 \label{eq:partition-function}
\end{equation}
A comment on the choice of eq.~(\ref{eq:observable}) is in order here.  
We have chosen the Gaussian functional 
$e^{-\frac{1}{2}\sum_i \mu^2 \phi_i^2}$ in eq.~(\ref{eq:observable}) 
for simplicity.
As one can easily see, it can be replaced with any
functional of $\phi$ as long as it gives a convergent and calculable
value as in eq.~(\ref{eq:C}).

Now let us introduce fermions.
After integrating out the fermions, we obtain
\begin{equation}
 \int \mathcal{D}\psi^*\, \mathcal{D}\psi\, \mathcal{D}\phi\,
  e^{-S_{\rm B} -S_{\rm F}}
 = \int \mathcal{D}\phi\, \sigma[D] e^{-S'},
\end{equation}
where $S_{\rm B}$ and $S_{\rm F}$ are the bosonic and fermionic part of the
action, $S'=S_{\rm B} -\ln |\det D|$ is an effective action with the
kernel $D$ of the fermion bilinear, and $\sigma[D]$ is the sign (or complex
phase) of $\det D$.\footnote{
Depending on the number of the fermions, the determinant
should be replaced by Pfaffian. }
The configurations for the ensemble average are generated using the
effective action $S'$ ignoring the sign factor $\sigma[D]$.
Let us denote the ensemble average over these configurations without the
sign factor as $\langle\ \cdot\ \rangle_0$.
The sign factor $\sigma[D]$ should be reweighted in the
measurement afterwards:
\begin{equation}
 \langle A \rangle
   = \frac{\int \mathcal{D}\phi\, A \sigma[D] e^{-S'}}
          {\int \mathcal{D}\phi\, \sigma[D] e^{-S'}}
   = \frac{\langle \sigma[D] A\rangle_0}{\langle \sigma[D] \rangle_0}.
 \label{eq:sign-reweight}
\end{equation}
To determine the normalization of partition function,
we also have to invert the effect of the sign factor
in addition to the contribution from the effective action $e^{-S'}$.
The analogue of eq.~(\ref{eq:observable}) becomes
$\langle \sigma[D]^{-1} 
  e^{S' -\frac{1}{2}\sum_i \mu^2 \phi_i^2}\rangle
 = C/\int\mathcal{D}\phi\, \sigma[D]e^{-S'}$, where the numerator gives again
the integration~(\ref{eq:C}).
Now we obtain the expression for the Witten index $w$:
\begin{equation}
 w=Z_{\rm P}= C\frac{\langle \sigma[D_{\rm P}] \rangle_{0,{\rm P}}}
  {\langle e^{S'_{\rm P}-\frac{1}{2}\sum_i \mu^2 \phi_i^2}\rangle_{0, {\rm P}}},
 \label{eq:partition-function2}
\end{equation}
provided all the fields are imposed the periodic boundary conditions as
indicated by the suffix~P.

Using pseudo fermion $\varphi$, we can rewrite
eq.~(\ref{eq:partition-function2}) as follow.
Our definition of the measure gives exactly
\begin{equation}
 \int \mathcal{D}\psi^*\mathcal{D}\psi 
  \exp \left[-\sum_{i,j}\psi^*_i D_{ij}\psi_j\right]
  = \int \mathcal{D}\varphi^{(1)}\mathcal{D}\,\varphi^{(2)} \,
    \sigma[D]\exp \left[-\sum_{i,j}\varphi_i^* (D^\dagger D)^{-1/2}_{ij} \varphi_j\right]
  = \det D,
\end{equation}
where $\varphi=\frac{1}{\sqrt{2}}(\varphi^{(1)}+i\varphi^{(2)})$.
Then, we obtain the expression for the index as
\begin{equation}
 w=Z_{\rm P}= C C_{\varphi}\frac{\langle \sigma[D_{\rm P}] \rangle_{0,{\rm P}}}
  {\langle e^{S''_{\rm P}
   -\frac{1}{2}\sum_i \mu^2 \phi_i^2
   -\sum_i\mu_\varphi^2 \varphi^*_i\varphi_i}\rangle_{0, {\rm P}}},
 \label{eq:partition-function3}
\end{equation}
where
\begin{align}
 S''&\equiv S_{\rm B}+\sum_{i,j}\varphi_i^* (D^\dagger D)^{-1/2}_{ij} \varphi_j,\\
 C_\varphi&\equiv \int \mathcal{D}\varphi^{(1)}\, \mathcal{D}\varphi^{(2)} \,
    e^{-\sum_i\,  \mu_\varphi^2 \varphi^*_i\varphi_i}.
  = \mu_\varphi^{-2N}.
\end{align}

It is interesting to see that there is a relation with the sign problem.
Eq.~(\ref{eq:sign-reweight}) implies 
that the phase quenched average of the sign factor
$\langle \sigma[D]\rangle_0$ is almost the partition function.
If this average is close to $0$, we cannot obtain reliable
expectation values numerically.  This is the sign problem and in fact it
occurs for supersymmetric quantum mechanics with spontaneously broken
supersymmetry \cite{KSS07}.
Note that if the Witten index is vanishing, the partition function is
vanishing so the expectation value becomes indefinite (or divergent)
with periodic boundary conditions. 
Therefore it seems reasonable to regard (not) having the sign problem
as an indication of the (non-)vanishing of the Witten index.\footnote{
The fact that $\langle \sigma[D]\rangle_{0,P}$ has a close relation to the
partition function and thus the Witten index
has been pointed out in \cite{KSS07}. 
See also \cite{Catterall:2008dv} for a related numerical result.}
 However, no justification
for this reasoning has been known, because no relation between the index
and $\langle \sigma[D_{\rm P}]\rangle_{0,{\rm P}}$ has been known so far.
Even if $\langle \sigma[D_{\rm P}]\rangle_{0,{\rm P}}$ is vanishing
within the small error in the simulation, say $0.01\pm 0.02$, without knowing 
the correct normalization, it could mean that the index is 
 $0.001\pm 0.002$ or $1.0\pm 2.0$.
Eq.~(\ref{eq:partition-function2}) makes a connection between these two
quantities and we can finally obtain the index from the sign factor.

One important disadvantage of the method is that it spoils the philosophy
of the important sampling method due to the factor $e^{+S'}$.
Therefore, the choice of the value of $\mu$ is important.  We have to 
chose it to have as large an overlap as possible between configurations and 
the operator we measure, namely $\exp[+S'-\frac{1}{2}\sum_i\mu^2 \phi_i]$.
The only adjustable parameter to obtain larger overlap is $\mu$.

Finally,
it is straightforward to extend our method 
to higher dimensional cases in principle.
The efficiency of the simulation, however, could be far from practical 
because of the overlap problem we have just mentioned above.

\section{Analytical check}
\label{sec:analytical}

The derivation of the path integral formulation from the operator
formulation assumes the continuum limit.  The lattice artifact may or
may not spoil the previous argument.
In this section, we discuss the effect of the exact supersymmetry on the
lattice.  We also give an explicit calculation of the index for the free
case in supersymmetric quantum mechanics.

Let us suppose that the lattice action has an exact supersymmetry $Q$,
like the one proposed in \cite{Sugino:2004qd}.
Then bosonic states and fermionic states make pairs even with finite
lattice spacings, 
except for states with $Q^\dagger Q=0$.  
At this stage, the situation is exactly the same as in the continuum.
The difference on the lattice is that we do not have $Q^\dagger$ as an
exact symmetry on the lattice.
However, supersymmetric vacuum must be annihilated by $Q$ and there is 
$Q$-exact Hamiltonian, so the conclusion does not 
change from the continuum case\footnote{Except for the overall sign. See the discussion below.}.  
Even with finite lattice spacing, 
if the index is a non-zero integer supersymmetry is not broken.
The advantage of exact invariance for the Witten index is also pointed 
out in \cite{Giedt:2004vb}.

A more rigorous argument is the following.
Using the fermion number operator $F$,
the Witten index is written as
\begin{equation}
 w=\Tr((-1)^F e^{-\beta H}).
 \label{eq:trace}
\end{equation}
For simplicity, we assume that the spectrum is discrete and the
index does not depend on $\beta\ (>0)$.
For a $Q$-exact action $S=Q\Lambda$ with $Q^2=0$, it is natural
to assume that the Hamiltonian $H$ in eq.~(\ref{eq:trace}) is also 
$Q$-exact:
\begin{equation}
 H=QJ_0.
\end{equation}
In fact, there is a natural way to define the $Q$-exact 
Hamiltonian~\cite{KSS07}.

Under the assumption of the $Q$-exactness of Hamiltonian, 
we can show that eq.~(\ref{eq:trace}) gives the index
even with the finite lattice spacing.
The Hamiltonian is complex in general in the
$Q$-exact formulation 
so we have to introduce left- and right-eigen states, which satisfy
\begin{align}
 H|L_i\rangle &= E_i|L_i\rangle,
 & \langle R_i|H &= \langle R_i|E_i, 
 & \langle R_i|L_j\rangle &= \delta_{ij}. 
\end{align}
The Witten index become
\begin{equation}
 w=\sum_i \langle R_i|e^{-\beta H} (-1)^F|L_i\rangle
\label{eq:index-rl}
\end{equation}
in this basis.
Since $Q$ commutes with $H$ (note that $H=\{Q, J_0\}$ in the operator
notation), $|L_i\rangle$ and $Q|L_i \rangle$ have the same eigen value $E_i$
for $H$.  These two states have opposite statistics so they do not
contribute to the index, except for the case $Q|L_i\rangle = 0$.
Similarly, $\langle R_i|$ and $\langle R_i|Q$ have the same eigen value
and thus do not contribute except for the case $\langle R_i|Q=0$.
Therefore, only $i$ such that 
$Q|L_i\rangle =0$ \emph{and} $\langle R_i|Q=0$ contribute to the summation 
in eq.~(\ref{eq:index-rl}).  These states have $E_i=0$, in fact:
\begin{equation}
 E_i
 =\langle R_i| H |L_i \rangle = \langle R_i|(QJ_0 + J_0Q)|L_i\rangle =0.
\end{equation}
Eventually, the index counts the number of the $E_i=0$ states with a weight
factor $(-1)^F$, which is exactly the same as in the continuum case.

This result holds even at finite lattice spacing,
because we have not assumed anything about the continuum
limit.\footnote{
Of course in order to guarantee that the continuum limit describes
the target theory correctly, 
we have to check the restoration of full supersymmetry other than $Q$.
}
In particular, the index on the lattice must be an integer.

We can check the argument above and our measure for the path integral
using a trivial example, i.e., the free supersymmetric quantum mechanics.
The index should be $1$ for the $Q$-exact lattice model.

The action of supersymmetric quantum mechanics with an exact $Q$
symmetry is
\begin{align}
 S
  &=-Q \sum_{k=0}^{N-1} \frac{1}{2}\psibar_k
  \left(F_k + \partial_+\phi_k +W'(\phi_k)\right) \\
  &=\sum_{k=0}^{N-1}\Bigl[
    \frac{1}{2}(\phi_{k+1}-\phi_k)^2 + \frac{1}{2}W'(\phi_k)^2
     +(\phi_{k+1}-\phi_k)W'(\phi_k) -\frac{1}{2}F_k^2 \nonumber\\
  &\qquad
    + \psibar_k(\psi_{k+1} -\psi_k) + W''(\phi_k)\psibar_k\psi_k
    \Bigr].
\label{eq:qexact-action}
\end{align}
Here, $\phi_k$ is a real scalar, $F_k$ is a bosonic auxiliary field,
and $\psi_k$ is a complex fermion,
$\partial_{+}\phi_k=\phi_{k+1}-\phi_k$, $W(\phi_k)$ is a potential and 
prime ($'$) denotes the derivative w.r.t.\ $\phi_k$. 
All fields are rescaled to be dimensionless.
$Q$-transformation, which is nilpotent, is:
\begin{align}
 Q\phi_k&=\psi_k, 
  & Q\psi_k &=0, \\
 Q\psibar_k &= F_k -\partial_{+}\phi_k -W'(\phi_k),
  & QF_k &= \partial_{+}\psi_k + W''(\phi_k)\psi_k.
\end{align}
In the following, we assume the free case:
\begin{equation}
 W(\phi_k)=\frac{1}{2}m \phi_k^2,
\end{equation}
where $m=am_{\rm phys}$ is the dimensionless mass.
The partition function with the periodic boundary conditions is
\begin{align}
 Z_{\rm P}
  &=\int_{\rm P} \mathcal{D}\phi\, \mathcal{D}F\,
 \mathcal{D}\psibar\, \mathcal{D}\psi\, e^{-S} 
 \nonumber \\
  &=\prod_{i=1}^{N-1}\left(
    \int_{-\infty}^{\infty}\frac{d\phi_i}{\sqrt{2\pi}}
    \int_{-\infty}^\infty\frac{dF_i}{\sqrt{2\pi}}
    \int d\psibar_i\, d\psi_i
    \right)
    \exp\left[
     -\frac{1}{2}\sum_{i,j}\phi_i B_{ij}\phi_j
     +\frac{1}{2}\sum_i F_i^2
     -\sum_{i,j}\psibar_i D_{ij} \psi_j
    \right] \nonumber \\
  &= (\det B)^{-\frac{1}{2}} \det D,
 \label{eq:Zp-qexact-free}
\end{align}
where
\begin{align}
 B_{ij}
  &=(1-m)\left( 2\delta_{ij}-\delta_{i-1,j}-\delta_{i+1,j} \right)
   +m^2\delta_{i,j},\\
 D_{ij}
  &=\delta_{i+1,j}-\delta_{i,j} +m\delta_{ij},
\end{align}
and $N$-periodicity is assumed for $B_{ij}$ and $D_{ij}$.
In obtaining eq.~(\ref{eq:Zp-qexact-free}), we made use of the replacement
$F\to iF$ in the integrand to make the integration over the auxiliary field $F$
convergent.  This gives unity with our measure 
and the contribution from the auxiliary field simply disappears.
Alternatively, one can regard that we have moved to the on-shell formulation
eliminating the auxiliary field, thus the integration should not contain
$F$ from the beginning which gives the same result.
It is straightforward to calculate the determinants $\det B$ and 
$\det D$.  Using the momentum representation, and keeping all the overall
normalization, we obtain
\begin{align}
 \det B
 &= (4-4m+m^2) m^2 \prod_{k=1}^{\frac{N}{2}-1} \left[
    (1-m)4\sin^2\frac{\pi k}{N} + m^2 \right]^2,\\
 \det D
 &= (-2+m)m \prod_{k=1}^{\frac{N}{2}-1} \left[
    (1-m)4\sin^2\frac{\pi k}{N} + m^2 \right]
\end{align}
for even $N$.
Assuming $m<2$, which is the case if the lattice spacing is small enough,
we obtain 
\begin{equation}
 Z_{\rm P}=-1 \qquad (\text{$N$: even}).
\end{equation}
For odd $N$, we obtain
\begin{align}
 \det B
 &= m^2 \prod_{k=1}^{\frac{N-1}{2}} \left[
    (1-m)4\sin^2\frac{\pi k}{N} + m^2 \right]^2,\\
 \det D
 &= m \prod_{k=1}^{\frac{N-1}{2}} \left[
    (1-m)4\sin^2\frac{\pi k}{N} + m^2 \right],
\end{align}
and thus 
\begin{equation}
 Z_{\rm P}=1 \qquad (\text{$N$: odd}).
\end{equation}
Except for the minus sign for even $N$, we obtain the correct Witten index
for the free case as expected.

What is the origin of the extra sign for even $N$\/?
It comes from $\det D$. 
In the derivation of the path integral, we have
\begin{equation}
 \langle \psi_{k+1}| \hat{\psi}^\dagger \hat{\psi} |\psi_l\rangle
 =\psi^*_{k+1}\psi_k \langle \psi_{k+1}| \psi_k \rangle,
\end{equation}
that is, $\psi_{k+1}^*$ (=$\psibar_{k+1}$) is always combined with $\psi_{k}$.
See eq.~(\ref{eq:sandwiched-by-psi}) in Appendix \ref{app:pathintegral}.
A deviation from this combination would give the derivative as in the
case in the kinetic term.
For the action we have used here, however, a natural combination 
is $\psibar_k \psi_k$.  The difference seems merely a lattice artifact
but it gives a factor $\prod_{k=0}^{N-1} e^{\frac{2\pi i}{N}k}=(-1)^{N+1}$
in the determinant, which explains the extra minus sign of $\det D$.
On the other hand, the bosonic part does not have such an artifact because
\begin{align}
 \langle q_{k+1}| \hat{q} |q_k \rangle
  &=  \langle q_{k+1}| q_k |q_k \rangle
   =   \langle q_{k+1}| q_{k+1} |q_k \rangle,
\end{align}
so it does not give any extra sign.

In the case of the naive action, the result is different.  The lattice
artifact survives and gives an extra contribution to the index.
Let us start with the following action:
\begin{align}
 S_{\rm naive}
 &=\sum_{i=0}^{N-1}\left[ 
  -\frac{1}{2}\phi_i(\phi_{i+1}+\phi_{i-1}-2\phi_i)
  +\frac{1}{2}m^2 \phi_i^2
  +\psibar_i(\psi_{i+1}-\psi_i) + m\psibar_i\psi_i
\right].
\end{align}
Therefore, we have
\begin{align}
 B_{jk}^{\rm naive}
  &= 2\delta_{jk}-\delta_{j,k+1}-\delta_{j+1,k} +m^2\delta_{jk}, \\
 D_{jk}^{\rm naive}
  &= \delta_{j+1,k}-\delta_{jk} +m\delta_{jk},
\end{align}
which gives
\begin{align}
 \frac{\det D^{\rm naive}}{(\det B^{\rm naive})^\frac{1}{2}}
 &= \begin{cases}
     \displaystyle
     \frac{-2+m}{\sqrt{4+m^2}} \prod_{k=1}^{\frac{N}{2}-1} 
     \frac{4(1-m)\sin^2\frac{k\pi}{N}+m^2}{4\sin^2\frac{k\pi}{N}+m^2}
    & (\text{$N$: even}) \\
     \displaystyle
     \prod_{k=1}^{\frac{N-1}{2}-1} 
     \frac{4(1-m)\sin^2\frac{k\pi}{N}+m^2}{4\sin^2\frac{k\pi}{N}+m^2}
    & (\text{$N$: odd})
    \end{cases}.
\end{align}
The contribution from the factors in front of the product
is $\mp 1$ in the continuum limit, where $m=am_{\rm phys}\to 0$.
For even $N$, the extra minus sign appears again.

The problem is the inside of the product.  Since there is no exact
relation between bosonic modes and fermionic modes, there is no exact
cancellation.  Each term has a structure
\begin{align}
 1-\mathcal{O}(m) &\sim 1-\mathcal{O}(1/N).
\end{align}
In the continuum limit, we obtain
\begin{align}
 \frac{\det D^{\rm naive}}{(\det B^{\rm naive})^\frac{1}{2}}
&\sim \prod_{k=1}^{N/2}\left(1-\frac{\alpha}{N}\right)
 \sim e^{-\frac{\alpha}{2}}
\end{align}
with a non-zero real number $\alpha$.  
It does not give the correct index even in the free case so
we cannot use the naive
action to measure the Witten index.

From the above argument, if the action has only $\mathcal{O}(a^2)$ or higher
lattice artifact (even without having any part of exact supersymmetry),
presumably we can use it to measure the Witten index.  For dimension $d\geq 2$,
the same argument allows only $\mathcal{O}(a^{d+1})$ and higher lattice
artifact, because the number of the product is $\sim N^d$.
In the case of higher dimensions, however, we will need a
fine tuning to obtain the supersymmetric continuum limit so the argument
here is too naive.

\section{Numerical result}
\label{sec:numerical}

In this section, we check our method using Monte Carlo simulation 
for supersymmetric quantum mechanics on the basis of the lattice formulation
(\ref{eq:qexact-action}).
We use the Hybrid Monte Carlo algorithm.  
See \cite{KSS07} for the implementation for this system\footnote{
In addition, we change the time step $\delta\tau$ 
in the leap-frog precess as follows.
In each trajectory we monitor the reversibility and if it is broken (it
occurs if $\delta \tau$ is not small enough) we restart that trajectory
using the same initial canonical momentum but smaller $\delta \tau$.   
Poor reversibilities are caused by very small eigen values of the Dirac operator.
}. 
We fix the physical volume $L_{\rm phys}=1$.  The lattice spacing
is $a=L_{\rm phys}/N=1/N$, where $N$ is the number of the lattice sites,
and the bare coefficients $\lambda_i$ in the potential 
are scaled as $\lambda_i^{\rm phys}N^{-i/2}$.
The parameters are summarized in table~\ref{tab:setlabel}.

\begin{table}
\hfil
 \begin{tabular}[t]{c|c}
  set & $\lambda_2^{\rm phys}$ \\
\hline
  2a   & 0.5 \\
  2b   & 1.0 \\
  2c   & 4.0 
 \end{tabular}
\hfil
 \begin{tabular}[t]{c|c|c}
  set & $\lambda_2^{\rm phys}$ & $\lambda_3^{\rm phys}$ \\
\hline
  3a   & 4.0  & 4.0 \\
  3b   & 4.0  & 16.0 \\
  3c   & 4.0  & 32.0 \\
  3d   & 2.0  & 16.0  
 \end{tabular}
\hfil
 \begin{tabular}[t]{c|c|c}
  set & $\lambda_2^{\rm phys}$ & $\lambda_4^{\rm phys}$ \\
\hline
  4a   & 1.0  & 1.0   \\
  4b   & 4.0  & 1.0   \\
  4c   & 4.0  & 4.0   
 \end{tabular}
 \caption{
 Set labels for the free case ($n=2$, left), the $n=3$ interaction (middle)
 and the $n=4$ interaction (right).
 We set $L_{\rm phys}=1$.
 }
 \label{tab:setlabel}
\end{table}

\subsection{Periodic case}

\begin{figure}
 \hfil
 \includegraphics{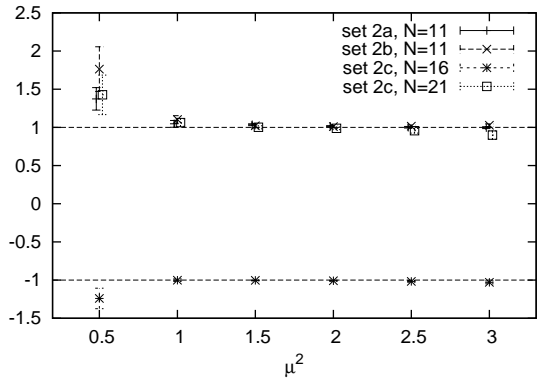}
 \caption{The index in the free case.  It gives $+1$ for even $N$ and $-1$
 for odd $N$ as discussed in the previous section.}
 \label{fig:n2pbc}
\end{figure}

The very first check is the free case with periodic boundary conditions.
There is only one bosonic vacuum and the index should be $1$.
We list the result in table~\ref{tab:n2pbc} and plot the index versus
$\mu^2$ in fig.~\ref{fig:n2pbc}.
We set $\mu^2=0.5, 1.0, 1.5, 2.0, 2.5, 3.0$.
The index from the simulation is in fact $1$ within the error (at worst
within 3 standard deviations)
for odd $N$ and $-1$ for even $N$.  
This result exactly agrees with what
we calculated in the previous section.

As table~\ref{tab:n2pbc} shows, 
large $\frac{1}{2}m_{\rm phys}=\lambda_2^{\rm phys}$ (set 2c)
requires less statistics, because the larger mass forces the scalar
field to stay around the origin, which makes larger overlap with the
quantity we measure. 
Note that large mass corresponds to low
temperature measured in unit of $m_{\rm phys}$, since it gives large
$L_{\rm phys} m_{\rm phys}$.

\begin{figure}
 \hfil
 \includegraphics{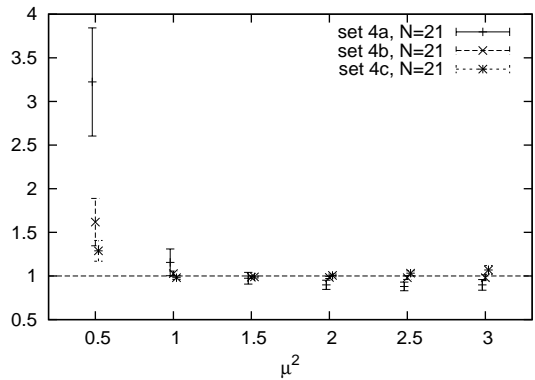}
 \caption{
 The index for the $n=4$ potential with various values of $\mu^2$.
 The dashed line represents the known result.
 }
 \label{fig:n4pbc}
\end{figure}

Next case is the supersymmetric case with interactions.  
We use the $n=4$ potential
\begin{equation}
 W(\phi)=\lambda_2 \phi^2 + \lambda_4 \phi^4.
\end{equation}
In this case, the index is known to be $1$ for $\lambda_2\neq 0$.
In table~\ref{tab:n4pbc}, we list the results for the index and
in fig.~\ref{fig:n4pbc} we plot the $\mu^2$ dependencies.
It reproduces the correct index within the error (except for 
$\mu^2=0.5$).
In particular, with a suitable choice of $\mu^2$ which gives small
enough error, we can easily 
identify the integer.  For some values of $\mu^2$, 0.5 for example,
the error is rather large and we cannot determine the integer value.
We expect, however, that if we used larger statistics it should converge 
to $1$.
For the strong coupling (set 4c), we found that the configuration contained 
a non-negligible amount of artifact configurations at larger lattice spacing.
We observed that for some bunch of configurations the artifact surface
term becomes as much as 10 times larger (or more) than the other part of
the action.
This is the same phenomena reported in \cite{Kastner:2008zc} 
in the 2-dimensional system.  However, such
artifact configurations disappear for smaller lattice
spacing.\footnote{
The author thanks Hiroshi Suzuki for pointing out the disappearing.
}
We observed this phenomena only for the $n=4$ case.

\begin{figure}
 \hfil
 \includegraphics{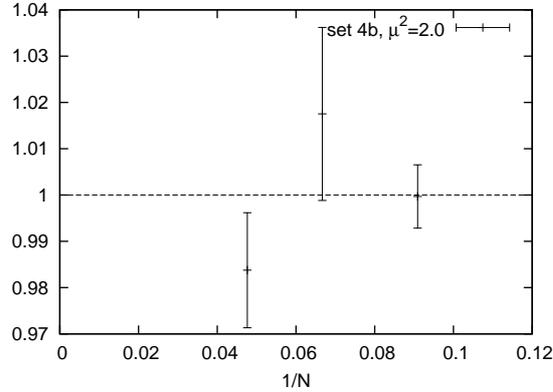}
 \caption{The index for the $n=4$ potential with different lattice spacings.}
 \label{fig:n4pbc_Ndep}
\end{figure}

In fig.~\ref{fig:n4pbc_Ndep}, we plot the dependence on the lattice
spacing $1/N$, which shows that the index we measured is
almost constant within the error against the lattice spacing.

\begin{figure}
 \hfil
 \includegraphics{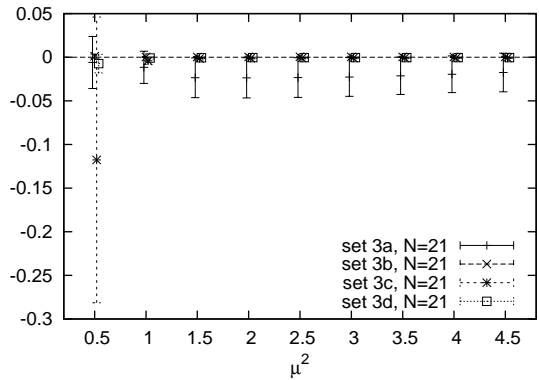}
 \caption{The index for the $n=3$ potential with various values of $\mu^2$.
 The dashed line represents the known result.}
 \label{fig:n3pbc}
\end{figure}

\begin{figure}
 \hfil
 \includegraphics{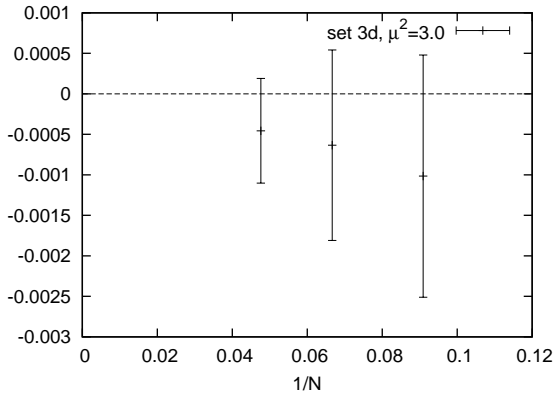}
 \caption{The index for the $n=3$ potential with different lattice spacings.}
 \label{fig:n3pbc_Ndep}
\end{figure}

Odd $n$ in the potential should give spontaneous breaking of 
supersymmetry.  The Witten index is known to be $0$.
We use $n=3$ potential:
\begin{equation}
 W(\phi)=\lambda_2 \phi^2 + \lambda_3 \phi^3.
\end{equation}
We list the results of the simulation in table~\ref{tab:n3pbc} and plot
some of them in fig.~{\ref{fig:n3pbc}} and \ref{fig:n3pbc_Ndep},
against $\mu^2$ and $1/N$, respectively.  
It gives in fact $0$ as expected.

A comment on the statistics is in order.
Since our measurement does not follow the philosophy of important
sampling, it has an overlap problem in general.
Therefore we need huge statistics.
We use large enough number of independent configurations $N_{\rm confs}$
to observe the behavior of the error 
$\sim N_{\rm confs}^{-1/2}$ at least for some values of $\mu^2$ 
which give small error.  See also the bottom of fig.~\ref{fig:errors}
for a typical behavior of the error.
For some values of $\mu^2$ which give large error (typically $\mu^2=0.5$),
the statistics are not large enough, since 
the error did not behave as $\sim (N_{\rm confs})^{-1/2}$.

In the measurement, we used configurations every 10 
trajectories for $n=2,\ 4$ case and 100 trajectories for $n=3$.
We used
$O(10^5)$--$O(10^6)$ configurations which may or may not be independent.
We estimated the error using the Jackknife method with several bin sizes. 
For $n=2,\ 4$, there were almost no 
autocorrelations in the measurement so that the bin size we
adopted is 1 for most sets of the parameter.
For $n=3$ case, however, the autocorrelation becomes 
longer and the bin size we adopted is 20--250.
This is because the sign 
$\sigma[D]$ does not change so frequently in the simulation.
The other parts of the observable like $S'$, which govern the
convergence of error, have almost no auto correlations.  Therefore,
even if the number of independent configurations is smaller than in the
$n=4$ case, which is governed by $\sigma[D]$, 
the number of the configuration is large enough 
to give the error $\sim N_{\rm confs}^{-1/2}$.

\begin{table}
 \hfil
 \begin{tabular}{c|c|r|c|l}
 set  &  $N$ & num. \hfil & $\mu^2$ & \hfil index \\
\hline
 2a  & 11 &  499000 & 2.0  & 1.014(9) \\
\hline
 2b  & 11 &  499000 & 2.5  & 1.017(15) \\
\hline
 2c  & 11 &  499000 & 1.0  & 0.9997(18) \\
     & 15 &   19900 & 1.5  & 0.990(7) \\
     & 16 &   99800 & 1.5  & -1.004(6) \\
     & 21 &  499000 & 2.0  & 0.989(17) \\
\end{tabular}

\caption{The free ($n$=2) case with periodic boundary conditions. We pick up
  $\mu^2$ which minimizes the error from $\mu^2=0.5, 1.0, 1.5, ...,
 3.0$.  The num.\ refers to the independent number of configurations
 after binning.}
\label{tab:n2pbc}
\end{table}

\begin{table}
 \hfil
 \begin{tabular}{c|c|r|c|l}
 set  &  $N$ & num. \hfil &  $\mu^2$ & \hfil index \\
\hline
 4a    & 11 & 199000 & 2.0  & 1.008(13) \\
       & 15 & 499000 & 2.0  & 1.018(19) \\
       & 21 & 999000 & 2.5  & 0.88(5) \\
\hline
 4b    & 11 &  99000 & 1.0  & 0.999(2)  \\
       & 15 &  49000 & 1.5  & 1.016(12) \\
       & 21 & 999000 & 2.0  & 0.984(12) \\
\hline 
 4c    & 11 & 99000 & 2.5  & 0.943(10)* \\
       & 15 & 99000 & 1.0  & 0.999(6)   \\
       & 21 & 499000 & 1.5  & 0.989(11) \\
 \end{tabular}

\caption{The $n=4$ case with periodic boundary conditions. 
 $\mu^2$ is chosen
 from 0.5,1.0,...,3.0 to minimize the error.
 Configurations for the set 4c with $N=11$ contain artifact configurations
 so the value (*) is not reliable.
}
\label{tab:n4pbc}
\end{table}

\begin{table}
 \hfil
 \begin{tabular}{c|c|r|c|l}
 set  &  $N$ & num. \hfil &  $\mu^2$ & \hfil index \\
\hline
 3a   & 11 & 1326 & 4.5  & -0.008(14) \\
      & 15 & 4995 & 2.0  &  0.002(5)  \\
      & 21 &  796 & 1.5  & -0.024(23) \\
\hline
 3b   & 11 & 1237 & 4.5  & 0.005(4)  \\
      & 15 & 1326 & 3.0  & -0.0018(21) \\
      & 21 & 3118 & 4.0  & 0.0004(7) \\
\hline
 3c   & 11 & 4950 & 4.5  & -0.001(3) \\
      & 15 & 2450 & 3.5  &  0.003(4) \\
      & 21 & 1237 & 3.0  & -0.0009(8) \\
\hline
 3d   & 11 & 3326 & 4.5  & -0.0010(14) \\
      & 15 & 1243 & 3.0  & -0.0006(12) \\
      & 21 &  660 & 3.0  & -0.0005(6)  \\
\end{tabular}
\caption{The $n=3$ case with periodic boundary condition.
 $\mu^2$ is chosen
 from 0.5,1.0,...,4.5 to minimize the error.
 }
\label{tab:n3pbc}
\end{table}

\subsection{Anti-Periodic case}
One may use anti-periodic boundary 
conditions for fermion in generating configurations.
Using a reweighting method, we can afterwards obtain the desired index.  
In this case what we measure is
\begin{equation}
 w= C\frac{\langle e^{S'_{\rm AP}}\sigma[D_{\rm P}]e^{-S'_{\rm P}}
 \rangle_{0,{\rm AP}}}
  {\langle e^{S'_{\rm A}
  -\frac{1}{2}\sum_i  \mu^2\phi_i^2}\rangle_{0,{\rm AP}}},
 \label{eq:w-from-AP}
\end{equation}
where we indicated the boundary condition, periodic (P) or anti-periodic
(AP) explicitly in the suffix.  Note the sign factor $\sigma$ must be
calculated from the periodic Dirac operator $D_{\rm P}$.
\begin{figure}
 \hfil
 \includegraphics{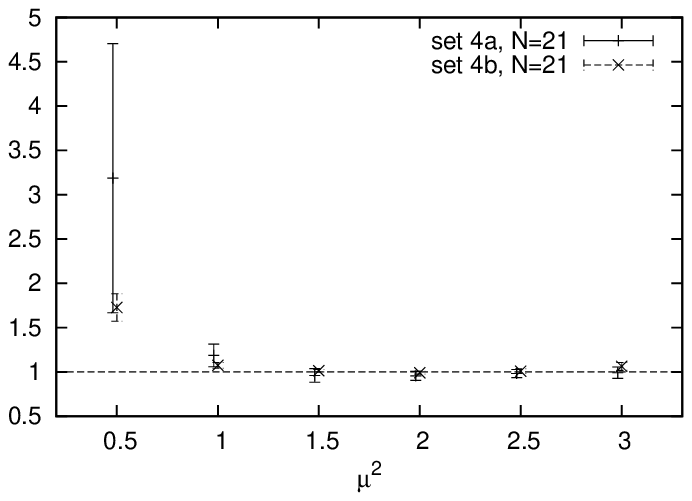}
 \caption{
The index for the $n=4$ potential with various values of $\mu^2$.
The configurations are generated with anti-periodic condition.
 The dashed line represents the known result.}
 \label{fig:n4apbc}
\end{figure}
\begin{figure}
 \hfil
 \includegraphics{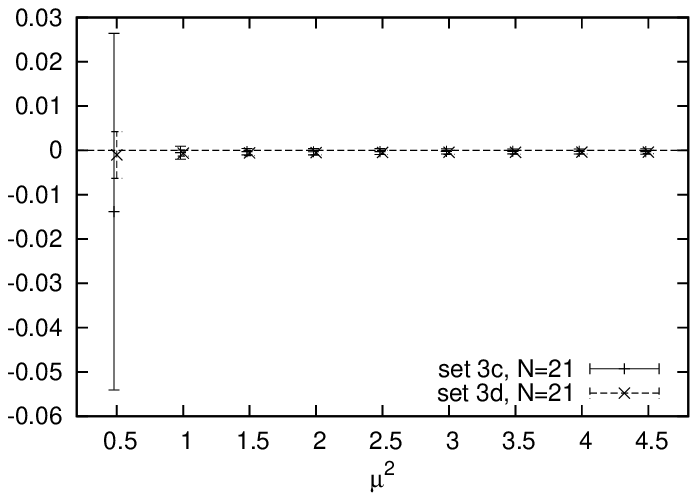}
 \caption{The index for the $n=3$ potential under various values of $\mu^2$.
The configurations are generated with anti-periodic condition.
 The dashed line represents the known result.}
 \label{fig:n3apbc}
\end{figure}

In fig.~\ref{fig:n4apbc} and \ref{fig:n3apbc}, we plot some of the
result against various values of $\mu^2$.
The detailed results are given in table~\ref{tab:apbc}.
It reproduces the desirable result as in the previous periodic case with
suitably chosen $\mu^2$.
Note that under the anti-periodic boundary conditions, supersymmetry is broken
because we impose different conditions for the boson and the fermion.
Even though, the $n=4$ case reproduces the correct index, although the
temperature of the system is rather low for set 4b thus 
the boundary condition may not give a large difference for the
configurations.  A rather non-trivial feature is found for the $n=3$ case.
The average of $\sigma[D_{\rm P}]$ is not necessarily  
vanishing but the index is
vanishing.  For example, $\langle \sigma[D_{\rm P}]\rangle_{0,{\rm AP}}$
for set 3d with $N=21$ is
$-0.018(7)$, which is slightly off from $0$, while the value of the index
from $\mu^2=4.5$ is
-0.0004(4).
The result shows that the reweighting in (\ref{eq:w-from-AP})
works perfectly.

\begin{table}
\hfil
 \begin{tabular}{c|c|r|c|l}
 set  &  $N$ & num. \hfil &  $\mu^2$ & \hfil index \\
\hline
 4a  & 11 & 499000 & 2.0  & 0.999(8)  \\
     & 15 & 999000 & 2.0  & 0.996(13) \\
     & 21 & 999000 & 2.5  & 0.98(5) \\
\hline
 4b  & 11 &  99000 & 1.0  & 0.9972(23) \\
     & 15 &  99000 & 1.5  & 0.990(8)  \\
     & 21 & 499000 & 1.5  & 1.012(14) 
\end{tabular}
\hfil
 \begin{tabular}{c|c|r|c|l}
 set  &  $N$ & num. \hfil &  $\mu^2$ & \hfil index \\
\hline
  3c & 11 & 1980 & 4.5  & 0.002(7) \\
     & 15 & 1960 & 4.5  & 0.000(3) \\
     & 21 & 6633 & 3.5  & -0.0002(6) \\
\hline
  3d & 11 & 1980 & 4.5  & 0.0007(27) \\
     & 15 & 1237 & 2.5  & 0.000(1)   \\
     & 21 & 1980 & 4.5  & -0.0004(4) 
 \end{tabular}

\caption{The index obtained from configurations with 
anti-periodic boundary conditions.
 $\mu^2$ is chosen
 from 0.5,1.0,...,3.0(set 4a,4b) or 0.5,1.0,...,4.5(set 3c, 3d) 
 to minimize the error.
}
\label{tab:apbc}

\end{table}

\subsection{The less important sampling method}
Since our method contradicts the philosophy of the important sampling,
we can prepare the configurations using ``less important'' sampling.
We replace the effective action $S'$ in generating the configuration 
by $(1-r)S'$, where $r$ is a real number.  
The ensemble average gives (assuming the large enough number of the
statistics)
\begin{equation}
 \langle A \rangle_r
  \equiv
  \frac{\int \mathcal{D} \phi\, A[\phi] e^{-(1-r)S'}}
 {\int \mathcal{D}\phi\, e^{-(1-r)S'}}.
\end{equation}
In order to obtain the original expectation value, 
we have to use the reweighting method.  For example,
\begin{equation}
 \langle A \rangle_0
 =\frac{\langle A e^{-rS'}\rangle_r}{\langle e^{-rS'}\rangle_r},
\end{equation}
and thus
\begin{equation}
 \langle A \rangle
 =\frac{\langle A \sigma[D] e^{-rS'}\rangle_r}{\langle \sigma[D]e^{-rS'}\rangle_r}.
\end{equation}
The Witten index becomes
\begin{equation}
 w = C 
    \frac{\langle \sigma[D_{\rm P}] e^{-rS'_{\rm P}}\rangle_{r,{\rm P}}}
         {\langle e^{(1-r)S'_{\rm P}
          -\frac{1}{2}\sum_i  \mu^2\phi_i^2 }
          \rangle_{r,{\rm P}}}
 \label{eq:index-less-important}
\end{equation}
for the configuration with the periodic boundary condition and 
\begin{equation}
 w = C 
  \frac{\langle \sigma[D_{\rm P}] e^{(1-r)S'_{\rm AP}}
  e^{-S'_{\rm P}}\rangle_{r,{\rm AP}}}
{\langle e^{(1-r)S'_{\rm AP}
-\frac{1}{2}\sum_i  \mu^2\phi_i^2 }\rangle_{r,{\rm AP}}}
\end{equation}
for the anti-periodic conditions.
Note that in gauge theories changing $r$ is nothing but changing the
bare gauge coupling $\beta$.

Now the advantage of the less important sampling is clear.
Let us set $r=\frac{1}{2}$ for simplicity and consider the periodic case.
In eq.~(\ref{eq:partition-function2}) the denominator has large fluctuations
while the numerator has much smaller fluctuations.
The behavior of the denominator requires large statistics and the unbalanced
magnitude of the fluctuation with that of the numerator means a poor efficiency.
In eq.~(\ref{eq:index-less-important}), the denominator is the square root
of that of eq.~(\ref{eq:partition-function2}), so the order of the
fluctuation becomes much smaller.  In addition, the numerator has the same
order of the fluctuation.  In total, we expect much more efficient measurement. 
\begin{figure}
 \hfil
 \includegraphics{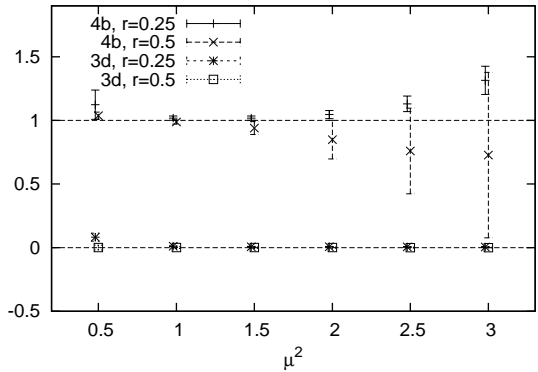}
 \caption{The index measured using the less important sampling for $N=21$.
 Set 4b should give $1$ and 3d should give $0$.}
 \label{fig:li_pbc}
\end{figure}

\begin{figure}
 \hfil
 \includegraphics{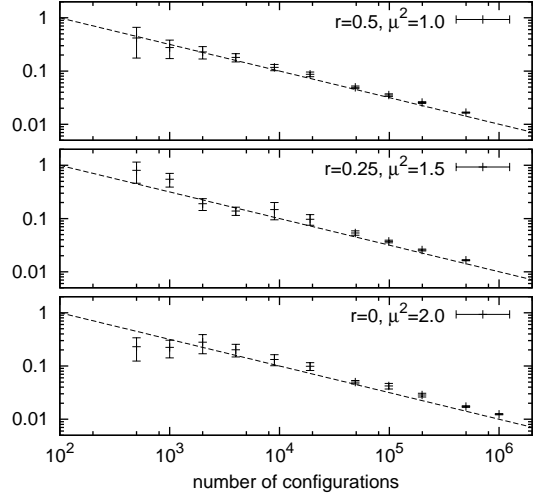}
 \caption{Behavior of the error versus number of independent
 configurations for different value of $r$.  They are for set
 4b, $N=21$ and $\mu^2$ with the smallest error.  Dashed lines are
 $10\times (N_{\rm confs})^{-1/2}$.}
 \label{fig:errors}
\end{figure}

In fig.~\ref{fig:li_pbc}, we show the result from the less important
sampling method for periodic-boundary conditions.  With a suitable choice
of $\mu^2$ which gives a small error, we can identify the correct index
as expected.
Table~\ref{tab:lipbc} summarizes the result based on the less important
sampling method
with $r=0.5$ and $r=0.25$.
To see the advantage, we plot the behavior of the error against the
number of configurations in fig.~\ref{fig:errors}, picking up set 4b.
Contrary to the expectation,
the magnitudes of the error are not so different from one another for 
larger statistics\footnote{
If we had carefully optimized the value of $\mu^2$, the magnitude of the
error for $r\neq 0$ might be significantly smaller than that for $r=0$.
}.
But the less important sampling case gives a more stable behavior.

An interesting byproduct is that it makes it easier to flip the sign
$\sigma[D_{\rm P}]$ for the $n=3$ case.  
The less important sampling method reduces the hight of the
potential barrier between the $\sigma[D_{\rm P}]=1$ region and the $-1$ region.
This effect gives less autocorrelation for the $n=3$ case compared with the
$r=0$ cases before (with either periodic or anti-periodic conditions).

\begin{table}
 \hfil
 \begin{tabular}{c|c|r|c|l}
 set  &  $N$ & num.\hfil &  $\mu^2$ & \hfil index \\
\hline
  4b & 11 &  99000 & 1.0  & 1.0049(21) \\
     & 15 &  99000 & 1.0  & 1.007(8)  \\
     & 21 & 499000 & 1.5  & 1.018(16) \\ 
\hline
  3d & 11 &    450 & 3.0  & 0.017(6) \\ 
     & 15 &    300 & 3.0  & -0.008(6) \\
     & 21 &    360 & 2.5  & 0.006(2) 
\end{tabular}
\hfil
 \begin{tabular}{c|c|r|c|l}
 set  &  $N$ & num.\hfil &  $\mu^2$ & \hfil index \\
 \hline
  4b & 11 &  99000 & 1.0  & 0.995(3) \\
     & 15 &  99000 & 1.0  & 1.010(11) \\
     & 21 & 499000 & 1.0  & 0.987(17) \\ 
 \hline
  3d & 11 &  12475 & 2.5  & -0.003(4) \\
     & 15 &   4975 & 3.0  & 0.005(4) \\
     & 21 &   9900 & 2.0  & 0.000(5) 
\end{tabular}

\caption{Results from the less important sampling method with $r=0.25$
 (left) and $r=0.5$ (right),
 under the periodic boundary conditions.
 $\mu^2$ is chosen
 from 0.5,1.0,...,3.0 to minimize the error.
}
\label{tab:lipbc}
\end{table}

\section{Conclusion and Discussion}

In this paper, we proposed a method for measuring the Witten index using
a lattice simulation.  The index is given by the average of the phase
of the fermion determinant and we gave a suitable normalization to
relate this average to the index.
Our requirement for the lattice model is that it
has at least one exact supersymmetry $Q$ with $Q^2=0$ and the action 
is $Q$-exact.   
We checked that the method reproduces the known Witten index for 
a simple supersymmetric quantum mechanics.   
The application to higher dimensional models is trivial in principle.

A disadvantage of the method is that it does not follow the philosophy of the 
important sampling.  Therefore it requires high statistics.  The situation
would be worse the more degrees of freedom the model has.  The efficiency
for the higher dimensional model would be less practical.
A less important sampling method may cure the situation to
some extent.

In addition, having more fields means having more parameters to be tuned
to achieve better efficiency.  The supersymmetric quantum mechanics we
numerically treated has only 1 parameter $\mu^2$, with our choice of
Gaussian regularization functional.  For the gauge theory, typically 
we have 3 such parameters; for scalar field, gauge field, and pseudo
fermion field.  One could even tune the regularization functional
to have more parameters.

For the gauge theory, however, it is difficult to give a suitable
gauge invariant regularization functional of the gauge fields.
We also have to normalize the gauge volume.
In the following, we discuss a possible treatment 
on the basis of the gauge fixing, though it contains some subtleties.
In the lattice simulation, we can fix the gauge in the measurement
setting as many link variables as possible to unity using the gauge
transformation.  The normalization constant $C$ should be obtained using
only the remaining degrees of freedom.  
Note that in the constraint system the relevant degrees of freedom 
which contribute to the path integral are the remaining ones after the
constraint is solved.  Setting the link variables unity using the gauge
transformation corresponds to solving the constraint.
An important subtlety here, which may spoil this gauge fixing argument,
is that it is not clear whether the gauge is completely fixed or not.  
Another subtlety is
the regularization functional for which throughout in this paper we used a 
Gaussian functional.  One can pick up the gauge field $A_\mu^a$ out of 
link variables and use the same Gaussian functional for $A_\mu^a$,
but this relation contains some lattice artifacts.
Therefore, it should give the correct normalization only after taking
continuum limit.  In this procedure the gauge fix is crucial since
otherwise non gauge invariant Gaussian functional cannot give a 
non-trivial expectation value.

Our method can be presumably applied to a lattice model with a $Q$-closed
term \cite{Hanada:2010kt},
which has an exact $Q$-invariance.  In this case we cannot exclude
contributions from $E>0$ states to the index.  We have to take
$L_{\rm phys}\to\infty$ limit for the temporal direction.

\subsection*{Acknowledgements}

The author thanks the Nishina Memorial Foundation for financial support.
He thanks A.~D'Adda for helpful discussions
and reading the manuscript carefully.
He also thanks to M.~Billo, M.~Hanada, R.~Lineros, H.~Matsufuru,
H.~Suzuki, M.~\"Unsal and U.~Wenger for useful comments and discussions.

\appendix
\section{Derivation of the path integral}
\label{app:pathintegral}

We give a short review of the derivation of the path integral.
\subsection{Bosons}
The bosonic Hamiltonian is
\begin{equation}
 \hat{H}(\hat{p},\hat{q}) = \frac{\hat{p}^2}{2m_{\rm QM}} + V(\hat{q}),
\end{equation}
where $m_{\rm QM}$ is a parameter with mass dimension 1 and $V(\hat{q})$ is a
potential.  

We calculate the amplitude
\begin{equation}
 \langle q_N(t=-i\tau = -iT) | q_0(t=-\tau=0) \rangle
 = \langle q_N | e^{-T\hat{H}} |q_0 \rangle,
\end{equation}
where $|q_0 \rangle$ and $|q_N \rangle$ are the initial and final state,
respectively.  The coordinate eigen state $|q\rangle$ is normalized as
\begin{equation}
 \langle q | q' \rangle = \delta(q-q')
\end{equation}
and satisfies the completeness relation
\begin{equation}
 \int_{-\infty}^{\infty} dq\, |q\rangle \langle q| = 1.
\end{equation}

We divide the period $T$ into $N$ small periods:
\begin{equation}
 T= aN.
\end{equation}
Inserting $N-1$ completeness relations, we obtain
\begin{align}
 \langle q_N | e^{-T\hat{H}} |q_0 \rangle
 &=  \left( \prod_{k=1}^{N-1}\int_{-\infty}^{\infty} dq_k \right)
      \langle q_N | e^{-a\hat{H}} |q_{N-1} \rangle
      \langle q_{N-1} | e^{-a\hat{H}} |q_{N-2} \rangle
      \cdots
      \langle q_{1} | e^{-a\hat{H}} |q_{0} \rangle,
\end{align}
and assuming $a$ is small, we obtain
\begin{align}
 \langle q_{k+1}|e^{-a\hat{H}} |q_k\rangle
 &= \langle q_{k+1} | (1 - a\hat{H}) |q_k \rangle \nonumber \\
 &= \int_{-\infty}^\infty \frac{dp_{k+1}}{2\pi}
    \exp\left[
     -a\frac{p_{k+1}^2}{2m_{\rm QM}} -aV\left(\frac{q_{k+1}+q_k}{2}\right)
     +ip_{k+1}(q_{k+1}-q_k)
    \right] \nonumber \\
 &= \left(\frac{m_{\rm QM}}{2\pi a}\right)^{\frac{1}{2}}
    \exp\left[ -a \left( 
     \frac{m_{\rm QM}}{2}\left(\frac{q_{k+1}-q_k}{2}\right)^2
     + V\left(\frac{q_{k+1}+q_k}{2}\right) \right)\right].
\end{align}
Here, we have assumed $\hat{H}$ is Weyl ordered.
To calculate $\langle q_{k+1}| \hat{H} |q_k\rangle$, 
we have also used the momentum eigen state $|p\rangle$ which satisfies
\begin{align}
 \langle p | q \rangle &= \frac{1}{2\pi}e^{-ipq}, &
 \int_{-\infty}^\infty dp |p\rangle \langle p| &=1.
\end{align}

In total, we obtain
\begin{align}
 \langle q_N | e^{-iT\hat{H}} |q_0 \rangle
 &= \left(\frac{m_{\rm QM}}{2\pi a}\right)^\frac{N}{2}
    \int_{-\infty}^\infty \left(\prod_{k=1}^{N-1}dq_k \right) \nonumber \\
 & \quad \times \exp\left[ -a \sum_{k=0}^{N-1} \left(
      \frac{m_{\rm QM}}{2} \left(\frac{q_{k+1}-q_k}{a}\right)^2 
      + V\left( \frac{q_{k+1}+q_k}{2} \right)
      \right)\right].   
\end{align}
In particular, the partition function becomes
\begin{align}
 \Tr (e^{-\hat{H}T}) 
   &= \int_{-\infty}^\infty dq_0 
      \langle q_N=q_0 | e^{-\hat{H}T} | q_0 \rangle \nonumber \\
   &=  \prod_{k=0}^{N-1} \left[
         \left(\frac{m_{\rm QM}}{2\pi a}\right)^{\frac{1}{2}}
          \int_{-\infty}^\infty dq_k \right] \nonumber \\
 & \quad \times \exp\left[ -a \sum_{k=0}^{N-1} \left(
      \frac{m_{\rm QM}}{2} \left(\frac{q_{k+1}-q_k}{a}\right)^2 
      + V\left( \frac{q_{k+1}+q_k}{2} \right)
      \right)\right],   
\end{align}
where the periodic boundary condition $q_N=q_0$ is assumed in the exponent.
Defining the 1-dimensional field $\phi_k=\sqrt{m_{\rm QM}}\, q_k$
(and suitable rescalings for parameters in the potential),
we obtain
\begin{align}
 \Tr (e^{-\hat{H}T}) 
   &= \prod_{k=0}^{N-1} \left[
         \left(\frac{1}{2\pi a}\right)^{\frac{1}{2}}
          \int_{-\infty}^\infty d\phi_k \right] \nonumber \\
 & \quad \times \exp\left[ -a \sum_{k=0}^{N-1} \left(
      \frac{1}{2} \left(\frac{\phi_{k+1}-\phi_k}{a}\right)^2 
      + V\left( \frac{\phi_{k+1}+\phi_k}{2} \right)
      \right)\right],   
\end{align}
where the first line provides the normalized measure.
The exponent goes to the (minus) action in $a\to 0$ limit.
Note that the above derivation gives one specific lattice action at
finite $a$, which may or may not be useful for the lattice simulation.

For a dimensionless lattice field
\begin{equation}
 \phi_k^{\rm lat} = a^{-\frac{1}{2}} \phi_k,
\end{equation}
the measure becomes
\begin{equation}
 \int \mathcal{D}\phi^{\rm lat}
 =
 \prod_{k=0}^{N-1} \left[
 \left(\frac{1}{2\pi}\right)^{\frac{1}{2}}
 \int_{-\infty}^\infty d\phi_k^{\rm lat} \right].
\end{equation}

\subsection{Fermions}

We use the following convention for the Grassmann integration:
\begin{align}
 \int d\psi^* d\psi\, \psi  \psi^* = +1.
\end{align}
The creation and annihilation operators $\hat{\psi}$ and 
$\hat{\psi}^\dagger$ satisfies the following anti-commutators:
\begin{align}
 \{\hat{\psi}, \hat{\psi}^\dagger\} &=1,
  &
 \{\hat{\psi},\hat{\psi}\}=\{\hat{\psi}^\dagger, \hat{\psi}^\dagger\} =0.
\end{align}
The Hilbert space is 2-dimensional and spanned by $|0\rangle$ and
$|1\rangle$ as usual:
\begin{align}
 \hat{\psi}|0\rangle &= 0,   & \hat{\psi}^\dagger |0\rangle &= |1\rangle, \\
 \langle 0 | 0 \rangle  &= \langle 1 | 1 \rangle =1,
 & \langle 0| 1\rangle &=0. 
\end{align}
The Hamiltonian for fermions is 
\begin{equation}
 \hat{H}(\hat{\psi}^\dagger, \hat{\psi})
  =\hat{\psi}^\dagger M \hat{\psi}.
\end{equation}

It is convenient to introduce the following coherent state:
\begin{equation}
 |\psi\rangle 
  \equiv e^{-\psi \hat{\psi}^\dagger}|0\rangle 
  = |0\rangle - \psi | 1 \rangle,
\end{equation}
where $\psi$ (without $\hat{\ }$) is a Grassmann odd c-number.
This state satisfies
\begin{equation}
 \hat{\psi} |\psi\rangle = \psi |\psi\rangle  \qquad (=\psi|0\rangle ).
\end{equation}
The orthogonality and the completeness relations are
\begin{align}
 \langle \psi|\psi'\rangle 
  &= 1 + \psi^* \psi' = e^{\psi^* \psi'} ,
 \\
 \int d\psi^* d\psi\, |\psi\rangle e^{-\psi^*\psi}\langle \psi|
  &= |0\rangle\langle 0| + |1\rangle\langle 1| =1.
\end{align}

Inserting $N-1$ complete sets and using
\begin{align}
 \langle \psi_{k+1}|e^{-a\hat{H}}|\psi_k\rangle
 &= \langle \psi_{k+1}|\left(1 -a H(\psi_{k+1}^*, \psi_k)\right)
    |\psi_k\rangle \nonumber \\
 &= \exp \left[
     -a H(\psi_{k+1}^*, \psi_k) + \psi_{k+1}^* \psi_k
    \right], \label{eq:sandwiched-by-psi}
\end{align}
we obtain
\begin{align}
 \langle \psi_N|e^{-\hat{H}T} | \psi_0\rangle
 &= \left( \prod_{k=1}^{N-1} \int d\psi_k^*\, d\psi_k\right)
    \exp\Biggl[
     -a\sum_{k=1}^{N-1}\left\{
     H(\psi_{k+1}^*, \psi_k)+\frac{1}{a}\psi_k^*(\psi_k-\psi_{k-1})
    \right\} \nonumber\\
  & \hspace{14em}
    +\psi_N^*\psi_{N-1} -a H(\psi_1^*,\psi_0)
    \Biggr].
\end{align}
The trace of an operator $\hat{A}$ in the coherent representation is
\begin{equation}
 \Tr(\hat{A})
  =-\int d\psi^*\, d\psi\, e^{\psi^*\psi} \langle \psi|\hat{A}|\psi \rangle
  = \langle 0|\hat{A}| 0 \rangle + \langle 1| \hat{A} | 1 \rangle .
\end{equation}
Then we obtain
\begin{align}
 \Tr(e^{-\hat{H}T})
  &=-\int d\psi_0^*\, d\psi_0\, e^{\psi_0^*\psi_0} 
    \langle \psi_0|e^{-\hat{H}T}|\psi_0 \rangle \\
  &= -\left( \prod_{k=0}^{N-1} \int d\psi_k^*\, d\psi_k\right)
    \exp\Biggl[
     -a\sum_{k=1}^{N-1}\left\{
     H(\psi_k^*, \psi_{k-1})+\frac{1}{a}\psi_k^*(\psi_k-\psi_{k-1})
    \right\} \nonumber\\
  & \hspace{14em}
    -aH(\psi_0^*,\psi_{N-1})
    +\psi_0^*\psi_{N-1} +\psi_0^*\psi_0
    \Biggr].
\end{align}
Replacing $\psi_0^* \rightarrow -\psi_0^*$ and imposing the anti-periodic 
condition $\psi_N \equiv -\psi_0$ and $\psi_N^* \equiv -\psi_0^*$,
we obtain
\begin{align}
 \Tr(e^{-\hat{H}T})
  &= \left( \prod_{k=0}^{N-1} \int d\psi_k^*\, d\psi_k\right)
    \exp\left[
     -a\sum_{k=0}^{N-1}\left\{
     H(\psi_{k+1}^*, \psi_{k})+\frac{1}{a}\psi_{k+1}^*(\psi_{k+1}-\psi_{k})
    \right\} 
 \right].
\end{align}
In the continuum limit $a\to 0$, the exponent gives the Euclidean action
with anti-periodic boundary conditions.

To conclude, the measure for the fermion field is
\begin{equation}
 \int \mathcal{D}\psi^*\mathcal{D}\psi 
  = \prod_{k=0}^{N-1} \int d\psi_k^*\,d\psi_k.
\end{equation}

\end{document}